\documentclass[aps,reprint,prb, superscriptaddress]{revtex4-1}
\usepackage[english]{babel}
\usepackage{graphicx}
\usepackage{dcolumn}
\usepackage{bm}
\usepackage{amssymb}
\usepackage{amsmath}
\usepackage{wasysym}
\usepackage{chngcntr}
\usepackage{textcomp}
\usepackage{hyperref}
\usepackage{setspace}

\usepackage{lipsum}

\begin{document}
\title{Resistive contribution in electrical switching experiments with antiferromagnets}
\author{Tristan Matalla-Wagner}
\email{tristan@physik.uni-bielefeld.de}
\author{Jan-Michael Schmalhorst}
\author{G\"unter Reiss}
\affiliation{Center for Spinelectronic Materials and Devices, Bielefeld University, Universit\"atsstra\ss e 25, D-33501 Bielefeld, Germany}
\author{Nobumichi Tamura}
\affiliation{Advanced Light Source, Lawrence Berkeley National Laboratory, Berkeley, CA, USA}
\author{Markus Meinert}
\email{markus.meinert@tu-darmstadt.de}
\affiliation{Department of Electrical Engineering and Information Technology, Technical University Darmstadt, D-64283 Darmstadt, Germany}

\date{\today}


\begin{abstract}
Recent research demonstrated the electrical switching of antiferromagnets via intrinsic spin-orbit torque or the spin Hall effect of an adjacent heavy metal layer. The electrical readout is typically realized by measuring the transverse anisotropic magnetoresistance at planar cross- or star-shaped devices with four or eight arms, respectively.
Depending on the material, the current density necessary to switch the magnetic state can be large, often close to the destruction threshold of the device. We demonstrate that the resulting electrical stress changes the film resistivity locally and thereby breaks the fourfold rotational symmetry of the conductor. This symmetry breaking due to film inhomogeneity produces signals, that resemble the anisotropic magnetoresistance and is experimentally seen as a ``saw-tooth''-shaped transverse resistivity. This artifact can persist over many repeats of the switching experiment and is not easily separable from the magnetic contribution. We discuss the origin of the artifact, elucidate the role of the film crystallinity, and propose approaches how to separate the resistive contribution from the magnetic contribution.
\end{abstract}

\maketitle

\counterwithout{equation}{section}

\section{Introduction}
\begin{figure}[t]
	\centering
	\includegraphics{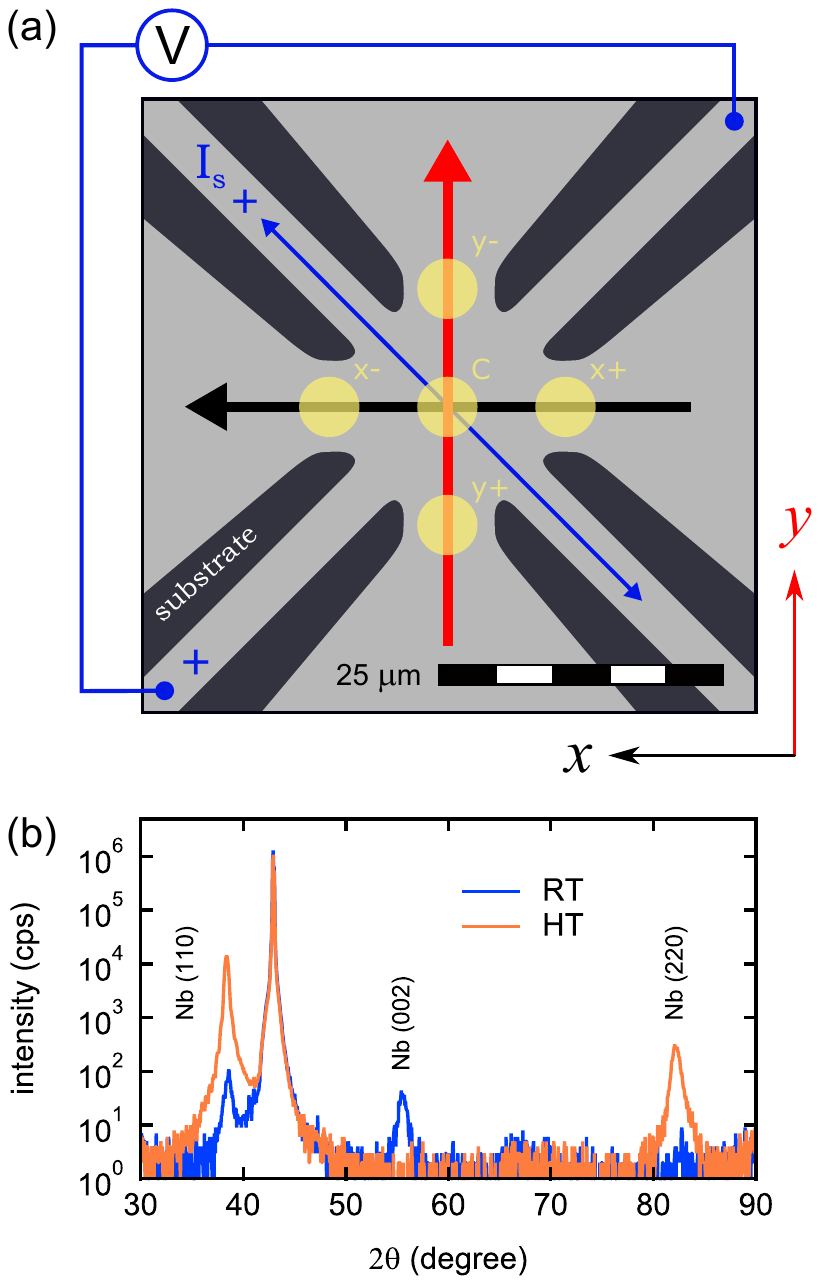}
	\caption[fig1]{\textbf{Sample preparation.}
		(a)~Schematic of the device. Current pulses can be applied to the pulse lines parallel to the $x$- or $y$-axis. The red and black arrows indicate the conventional current direction for positive polarity. An AC sensing current $I_\text{s}$ is applied to a probe line and the voltage $V$ is measured perpendicular to $I_\text{s}$. The pulse and probe lines have a width of $8\,\text{\textmu m}$ and $4\,\text{\textmu m}$, respectively. Yellow circles show the positions of the x-ray spot for \textmu XRD measurements and the nomenclature for the positions.
		(b)~XRD measurements of unpatterned Nb films grown at room-temperature (RT) and $400\,^\circ\text{C}$ (high-temperature, HT) directly after deposition.
	}
	\label{fig:1}
\end{figure}

The possibility to switch the magnetic order of antiferromagnets (AFM) with electric currents has been predicted about five years ago \cite{Zelezny2014}. It was theoretically shown that materials of certain symmetry can exhibit staggered nonequilibrium spin polarization. If these coincide with the crystallographic positions of the staggered magnetic moments, a net torque on the antiferromagnetic order arises, which enables its current-driven reorientation. While the effect was predicted for Mn$_2$Au, its first realization was demonstrated with CuMnAs \cite{Wadley2016}. Later, the N\'eel-order switching was also shown for Mn$_2$Au films \cite{Meinert2018, Bodnar2018, Zhou2018}. A greater variety of AFM can be switched using heterostructures that utilize the spin Hall effect (SHE) in an adjacent heavy metal (HM) like Pt to exert an antidamping torque on the N\'eel-order \cite{Chen2018, Moriyama2018, Gray2018, Baldrati2018, Zhou2019, Dunz2019}. The reorientation can be realized with cross-shaped planar devices, where electrical pulses driven through the orthogonal lines allow for a reproducible switching of the N\'eel-order by 90$^\circ$. In AFM/HM heterostructures there are two possible readout mechanisms with identical symmetry. The spin-polarized current in the HM is partially absorbed by the AFM, which depends on the relative orientation of the polarization and the magnetic moments in the AFM. This gives rise to the spin Hall magnetoresistance (SMR). The second mechanism is the anisotropic magnetoresistance (AMR). In both cases, typically $\rho_{xx} > \rho_{yy}$ for the current flowing parallel or perpendicular to the magnetic moments is found. In single layer switching experiments, only the AMR is available for electrical readout. To observe the rather small resistivity change associated with the AMR or SMR, it is favorable to measure in transverse geometry. For a sense current density $j$, the associated transverse electric field is given by $E_\perp = j\,(\rho_{xx} - \rho_{yy}) \sin{(2\varphi)} / 2$, where $\varphi$ is the angle between sense current density and magnetic moments. The experiment is performed by recording the transverse voltage $V_\perp = R_\perp I_\mathrm{s}$ with a fixed sense current $I_\mathrm{s}$ in a device as sketched in Fig.~\ref{fig:1}\,(a). In this type of experiments, Cheng \textit{et al.} identified a ``saw-tooth''-shaped contribution in $R_\perp$ measurements arising from the Pt layer for large current density \cite{Cheng2019}, which is of nonmagnetic origin. It was found to be larger than the magnetic contribution to the transverse electrical response in some cases.

In the present manuscript, we investigate the origin of this nonmagnetic contribution to the electrical response of the star-shaped structures. We chose paramagnetic Nb films as a model system, because here no magnetic contribution is present. Two different types of films are compared, grown either at room temperature (polycrystalline) or at high temperature (epitaxial). We analyze the $R_\perp$ response to current pulses and inspect the star-shaped devices \textit{in operando} by x-ray microdiffraction (\textmu XRD) and scanning electron microscopy (SEM). Our findings are consistent with the results presented in Ref.~\cite{Cheng2019}, where a change of $R_\perp$ with the planar Hall effect symmetry is observed in Pt films. We interpret this observation as a local variation of the film resistivity which gives rise to a $R_\perp$ change $\Delta R_\perp$ with alternating sign for the different pulsing directions. Calculations using a finite-element-method (FEM) show that the effect size is in quantitative agreement with this explanation. Furthermore we see that the sign of $\Delta R_\perp$ can change within one experiment highlighting that at least two processes of nonmagnetic origin can be present in such experiments, related to locally decreasing or increasing resistivities.

\section{Sample characteristics and experimental setup}

Samples of MgO (001) / Nb 25\,nm / Si 2\,nm were grown by magnetron sputtering at room temperature (RT) and at $400\,^\circ\text{C}$ (high-temperature, HT). The Si capping prevents oxidation of the films. The film quality is characterized by x-ray diffraction (XRD). The RT sample shows no preferred growth direction, whereas the HT sample grows in (110) direction, see Fig.~\ref{fig:1}\,(b). The resistivities are $\rho_\text{RT} = 27.4\,\text{\textmu}\Omega\text{cm}$ and $\rho_\text{HT} = 23.5\,\text{\textmu}\Omega\text{cm}$, determined by four-point measurements on the films. Finally, the films are patterned to star-shaped devices as depicted in Fig.~\ref{fig:1}\,(a) by photolithography and wire-bonded into IC packages. On each sample, half of the devices have their pulse lines $x$ ($y$) aligned parallel to the [100] ([010]) direction of the substrate, the other half of the devices are rotated by $45^\circ$.

\begin{figure}[b]
	\centering
	\includegraphics{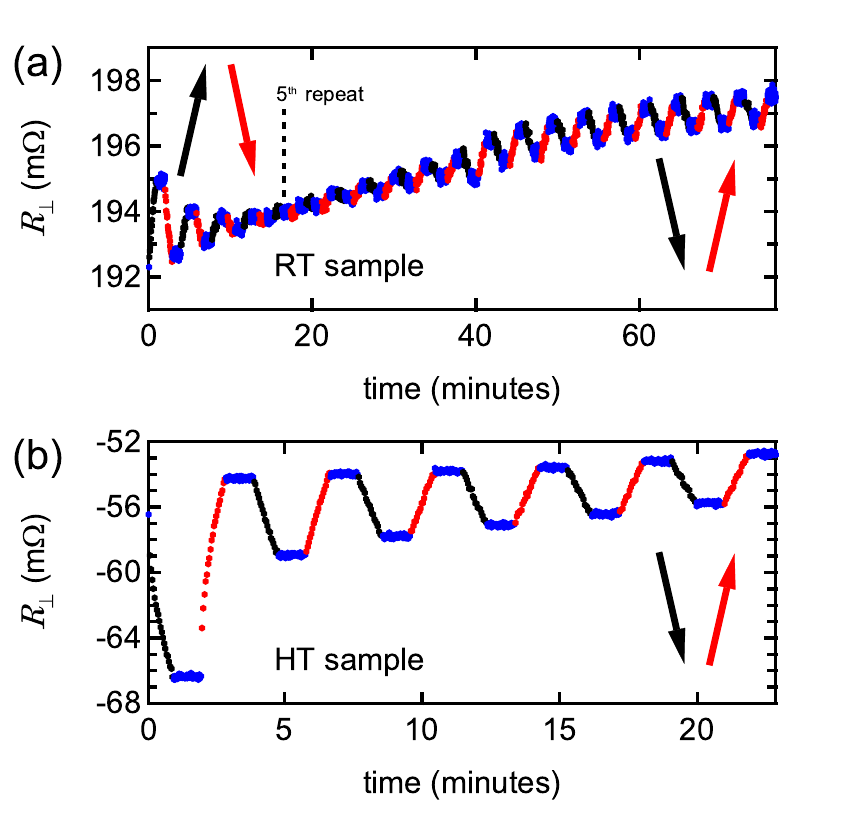}
	\caption[fig2]{\textbf{Pulsing experiment.}
		(a)~Pulses with $j = 7.5 \times 10^{11}\,\text{A}/\text{m}^2$ and $\Delta t = 10\,\text{\textmu s}$ are applied to a device on the RT sample with $x \parallel  \text{MgO}[100]$. The colors of the data points match the arrows in Fig.~\ref{fig:1}\,(a) to identify the pulsing and the relaxation phase. The arrows here accentuate the sign of $d R_\perp / dt$ that correlates with a certain pulsing direction. This sign inverts after the fifth repeat.
		(b)~A similar measurement performed at the HT sample with $j = 11.0 \times 10^{11}\,\text{A}/\text{m}^2$ and $\Delta t = 2\,\text{\textmu s}$.
	}
	\label{fig:2}
\end{figure}

\begin{figure*}[t]
	\centering
	\includegraphics{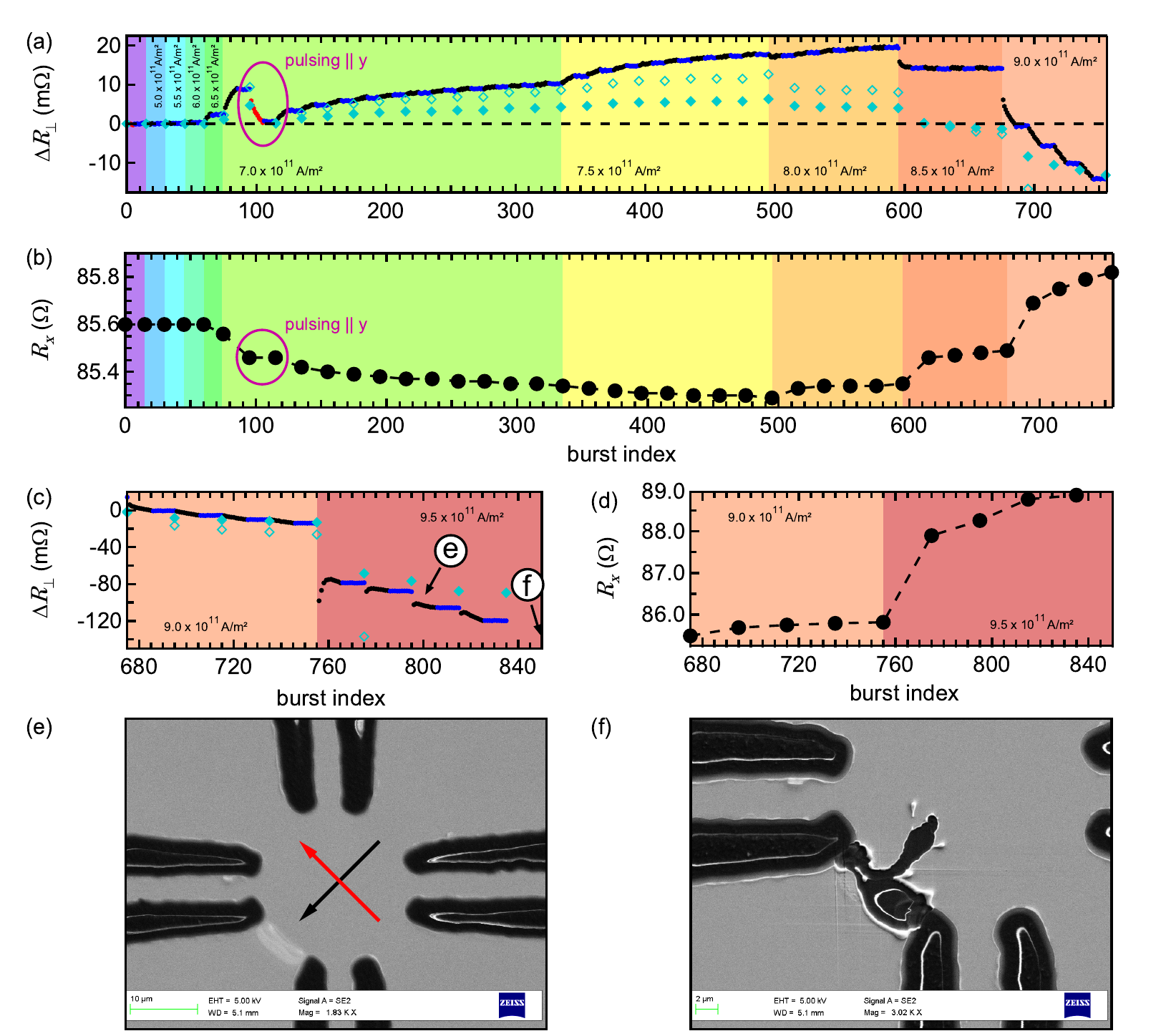}
	\caption[fig3]{\textbf{SEM experiment.}
		Pulses using $\Delta t = 10\,\text{\textmu s}$ are applied to a device on the RT sample with $x \parallel  \text{MgO}[110]$.
		(a),~(c)~$R_\perp$ response to bursts of different $j$. The colors of the data points match the arrows (red, black) in Fig.~\ref{fig:1}\,(a) to identify the pulsing direction and the relaxation phase (blue). Filled diamonds in (a) and (c) show the expected $\Delta R_\perp$ calculated from $R_x$ in (b) and (d) by FEM simulations considering identical resistivity changes in opposing constrictions. The open diamonds are scaled by a factor of two.
		(a)-(d)~The background of the plots highlights which magnitude of $j$ is applied. Encircled letters in (c) mark the points at which the SEM images in (e) and (f) have been recorded.
		(b),~(d)~$R_x$ is measured prior to a pulsing-relaxation sequence.
		(e),~(f)~SEM images of the device after the second (e) and fifth (f) repeat with $ j \le 9.5 \times 10^{11}\,\text{A}/\text{m}^2 $. The arrows in (e) indicate how the device was contacted (cf. Fig.~\ref{fig:1}\,(a)) with the negative polarity for pulsing in $x$ direction in the lower left.
	}
	\label{fig:3}
\end{figure*}

The experimental procedure for the electrical experiments is similar to our experiment on N\'eel-order switching in CuMnAs, see Ref.~\cite{Matalla-Wagner2019}. We apply $n$ bursts of current pulses to our structure in one of two orthogonal directions $x$ and $y$ (cf. Fig.~\ref{fig:1}\,(a)). The pulses have a length of $\Delta t = 1 - 10\,\text{\textmu s}$ and a duty cycle of $10^{-2}$. The number of pulses per burst is not constant in this work but typically a burst consists of around 100 pulses. After each burst, we measure $R_\perp$ with a lock-in amplifier. We also observe the evolution of $R_\perp$ after the burst sequence during a relaxation-phase. The pulse-line resistances $R_x$ and $R_y$ are measured prior to the burst sequence with a Keithley $2000$ multimeter to set the voltage output of the arbitrary waveform generator accordingly. The experiments are additionally performed with simultaneous \textmu XRD measurements or SEM imaging. The \textmu XRD investigation was conducted at beamline $12.3.2$ of the Advanced Light Source (ALS), Berkeley, California, USA \cite{Kunz2009}. We measure diffraction patterns with a PILATUS 1M detector and an integration time of $10 - 30\,\text{s}$ for the beam located on the constrictions of each pulseline and in the center of the device, as highlighted by the yellow areas in Fig.~\ref{fig:1}\,(a). The spot size is approximately 5\,\mbox{\textmu m} and the energy is tuned to match the diffraction conditions.

\section{Results}

A pulsing experiment performed with Nb thin films is presented in Fig.~\ref{fig:2}. Panel (a) shows the change of $R_\perp$ for the RT sample with a current density of $ j = 7.5 \times 10^{11}\,\text{A}/\text{m}^2 $ applied alternating to the $x$ and $y$ pulseline. The respective relaxation is observed for $T_\text{relax} = 60\,\text{s}$. We observe a significant change of $R_\perp$ during pulsing, whereas $R_\perp$ remains stable during the relaxation. Initially, pulsing along $x$ increases $R_\perp$ while pulsing along $y$ decreases $R_\perp$. The amplitude of this \glqq saw-tooth\grqq~signal decreases with each repeat and vanishes around the fifth repeat. The following repeats show an inverted sign of the response with virtually constant amplitude. An overall trend towards higher values of $R_\perp$ is present which is most likely related to asymmetries arising from imperfect lithography. The polarity of the pulse current has no impact on the signal (not shown). Panel (b) of Fig.~\ref{fig:2} shows an equivalent measurement for the HT sample. However, a higher current density of $j = 11.0 \times 10^{11}\,\text{A}/\text{m}^2$ had to be applied to observe a transverse electrical response of similar magnitude. From a current density variation we find, that the HT sample only shows one sign of transverse resistance change. The degradation of the transverse resistance amplitude slows down for successive repeats. We observe no qualitative difference in behavior between devices with $x \parallel  \text{MgO}[100]$ or $x \parallel  \text{MgO}[110]$.

In the following, we investigate a fresh device on the RT sample and take SEM images after each pulsing experiment with the device grounded while taking the image. Here, we primarily pulse along $x$ and use pulses in $y$ direction only to demonstrate that the sign of $\Delta R_\perp$ depends on the pulsing direction as before. The $R_\perp$ response of the system and the values of $R_x$ taken intermittently are plotted in Fig.~\ref{fig:3}\,(a) and (b), respectively. For $ j \le 6.0 \times 10^{11}\,\text{A}/\text{m}^2 $ we neither see a transverse response nor a change of $R_x$. Starting with $ j = 6.5 \times 10^{11}\,\text{A}/\text{m}^2 $ we observe a response with $\Delta R_\perp > 0$ and a decrease of $R_x$. Further increasing $j$, enhances the transverse response and the resistance change $\Delta R_x$ alike. Pulsing in $y$ direction changes the sign of $\Delta R_\perp$ and has no impact on $R_x$. For successive bursts in the same direction, $\Delta R_\perp$ and $\Delta R_x$ seem to saturate until $j$ is increased again. At $ j = 8.0 \times 10^{11}\,\text{A}/\text{m}^2 $, we observe a crossover between reducing and increasing $R_x$, where the change of the transverse response also changes sign. This change in sign becomes more apparent for even higher $j$ until a response similar to Fig.~\ref{fig:2}\,(b) is obtained for $ j = 9.0 \times 10^{11}\,\text{A}/\text{m}^2 $. The effect size greatly enhances for $ j = 9.5 \times 10^{11}\,\text{A}/\text{m}^2 $ as seen in Fig.~\ref{fig:3}\,(c) and (d). For this current density, we observe that the transverse response intermittently has a positive sign, while the overall trend is negative. $R_x$ increases with each successive experiment until the device breaks. We were not able to resolve any change of the device in SEM images for $j \le 9.0 \times 10^{11}\,\text{A}/\text{m}^2 $. However, from the first sequence at  $ j = 9.5 \times 10^{11}\,\text{A}/\text{m}^2 $ until the destruction of the device we see clear changes arising at the lower left constriction of the structure. An exemplary image taken after the second repeat with highest $ j $ is shown in Fig.~\ref{fig:3}\,(e). With the first burst of the fifth repeat the device breaks (cf. Fig.~\ref{fig:3}\,(f)) with droplets of material moving in electron flow direction.

\begin{figure}[t]
	\centering
	\includegraphics{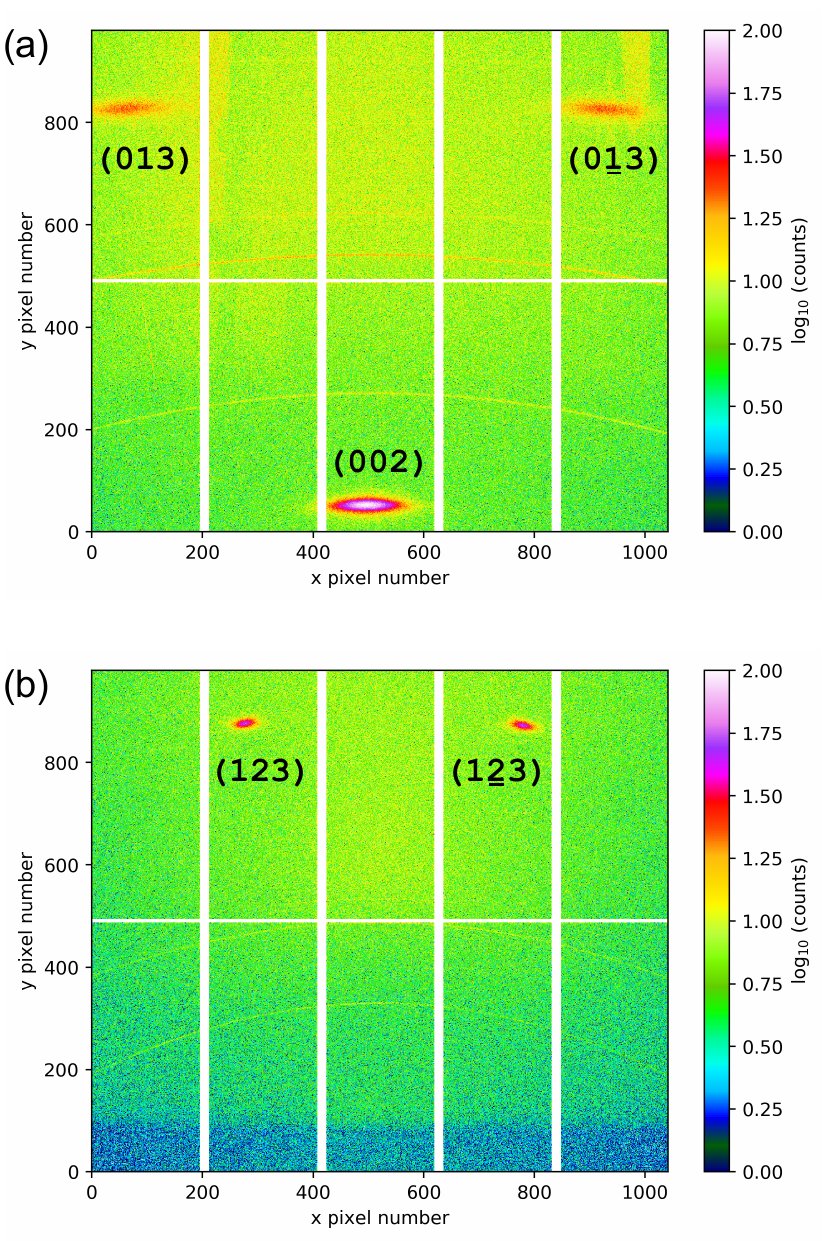}
	\caption[fig4]{\textbf{\textmu XRD experiment.}
		Exemplary 2D diffractograms measured with monochromatic x-rays of energy $h\nu$. The reflexes result from the Nb film while the rings are due to the Al bonding wires.
		(a)~RT sample measured with $h\nu = 11.2\,\text{keV} $.
		(b)~HT sample measured with $ h\nu = 12.6\,\text{keV} $. The low-intensity area at the bottom is caused by shadowing from the IC package.
	}
	\label{fig:4}
\end{figure}

\begin{figure}[h!]
	\centering
	\includegraphics{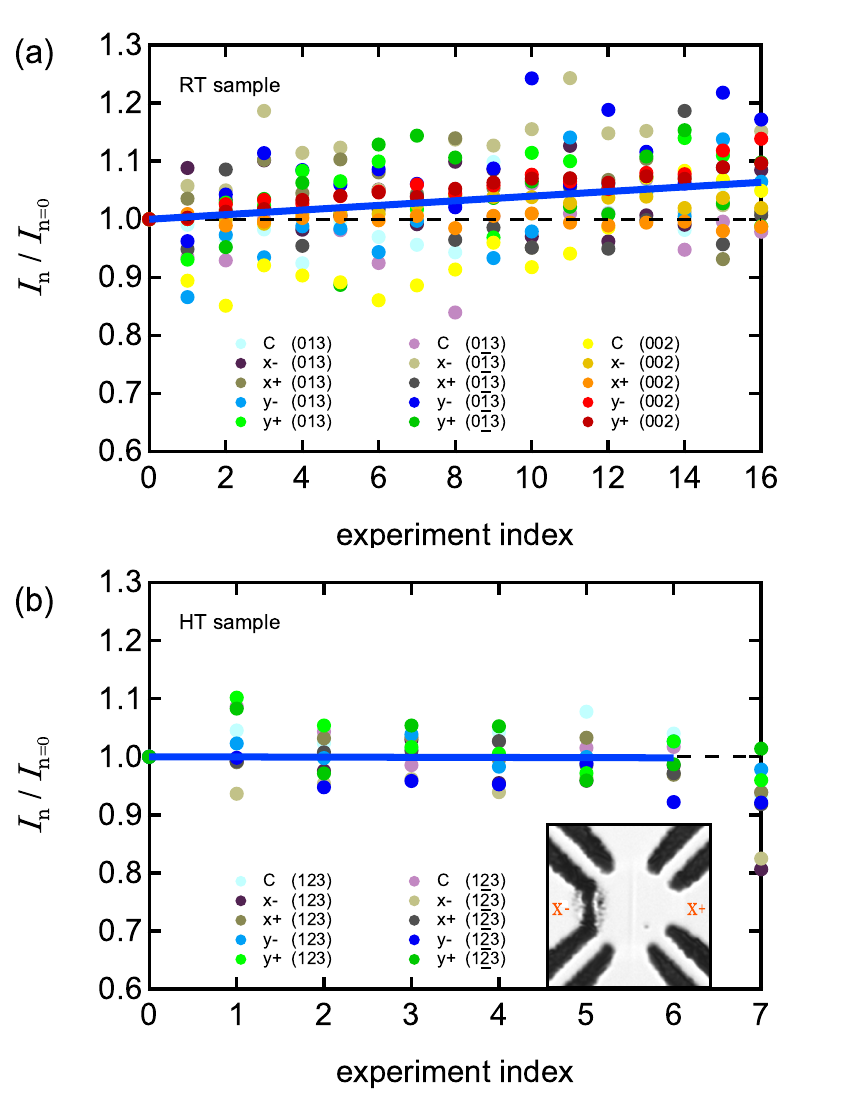}
	\caption[fig5]{\textbf{\textmu XRD evaluation.}
		Plots of the normalized peak intensities $I_n$ divided by their values prior to the pulsing experiment for (a) the RT and (b) the HT sample. The experiment index increases with every pulsing-relaxation sequence. 2D diffractograms are taken and evaluated for each peak (cf. Fig.~\ref{fig:4}) and every position on the sample (cf. Fig.~\ref{fig:1}\,(a)). The dashed black lines represent $I_n / I_{n=0} \equiv 1$. Linear fits of the data are shown as blue lines. The fit in (b) is applied for experimental index $n=0\dots 6$. The inset shows a micrograph of the broken device after $n=7$.
	}
	\label{fig:5}
\end{figure}

We identified a correlation between $R_x$ and $R_\perp$ where the physical origin of this resistance change needs to be clarified. In the following, we attempt to investigate whether local annealing due to the Joule heating may contribute to the observed behavior. To this end, we perform \textmu XRD observations of selected x-ray diffraction peaks while performing pulsing experiments. Exemplarily, we show 2D diffractograms in Fig.~\ref{fig:4}. The positions of the x-ray focus are shown in Fig.~\ref{fig:1}\,(a). During the experiment, we do not observe any changes of the peak positions. To determine the crystallinity of the samples, we define regions of interest (ROI) around the peak positions and calculate the integrated intensity $I_{n\mathrm{, raw}} = \sum^\text{pixel} i_\text{pixel}$, with $i_\text{pixel}$ being the counts of each pixel and $n$ being the experiment index. The x-ray intensity of the beamline varies slightly over time, so we normalize $I_n$ to the background signal $I_\text{bg}$ which is calculated from an ROI that contains only diffuse scattering and fluorescence signals. In Fig.~\ref{fig:5}, we plot the normalized peak intensity $I_n = I_{n\mathrm{, raw}} / I_\text{bg}$ divided by its initial value $I_{n=0}$ against the experiment index, which is essentially a number raised with each pulsing-relaxation sequence containing no information about $j$ or the pulsing direction. (a) and (b) show the data measured at the RT and HT sample, respectively. We repeat pulsing experiments either until the sign of the response inverts or the device breaks. Unfortunately, the signal-to-noise ratio is rather unsatisfying despite the integration times of up to $30\,\mathrm{s}$ per image. Instead of looking at the individual positions, we collect all datasets and apply a linear regression $y = y_0 + m \cdot n$ with fixed $y_0=1$ to extract the trend of peak intensity under repeated electrical pulsing. In Fig.~\ref{fig:5}, these regressions are visualized by the blue lines. We find a significantly positive slope of $m_\text{RT} = 4.0\permil \pm 0.5\permil$ for the device on the RT sample. The HT sample was destroyed during the experiment with index 7 where pulses were applied in $x$ direction. It can be seen in Fig.~\ref{fig:5}\,(c) that the peak intensity measured at the $x-$ position after destruction is significantly reduced compared to the other positions. An optical micrograph verifies that indeed the $x-$ constriction is the one which broke during the experiment. Linear regression with $n=0\dots 6$ gives $m_\text{HT} = -0.3\permil \pm 1.3\permil$, i.e. no significant change in peak intensity is observed. Thus we conclude that the crystallinity of the RT sample is enhanced due to the electrical pulsing. For the HT sample, no enhancement of the crystallinity is observed.

\section{Discussion}

When a sense current $I_\text{s}$ is passed through a material, the appearance of a transverse electric field is generally due to a breaking of the mirror symmetry of the resistivity about the current direction. For the AMR and SMR, this is due to $\rho_{xx} \neq \rho_{yy}$ and the current being neither parallel nor perpendicular to the magnetic moments. As we have shown in this study, one can observe a transverse voltage change induced by electrical pulses that largely resembles the electrical switching of the N\'eel-order in antiferromagnets in devices that are nonmagnetic. Although the amplitude of this nonmagnetic response eventually reduces with successive pulsing, the decay can be marginal and looks like a practically constant amplitude over dozens of repeats.

\begin{figure}[t]
	\centering
	\includegraphics{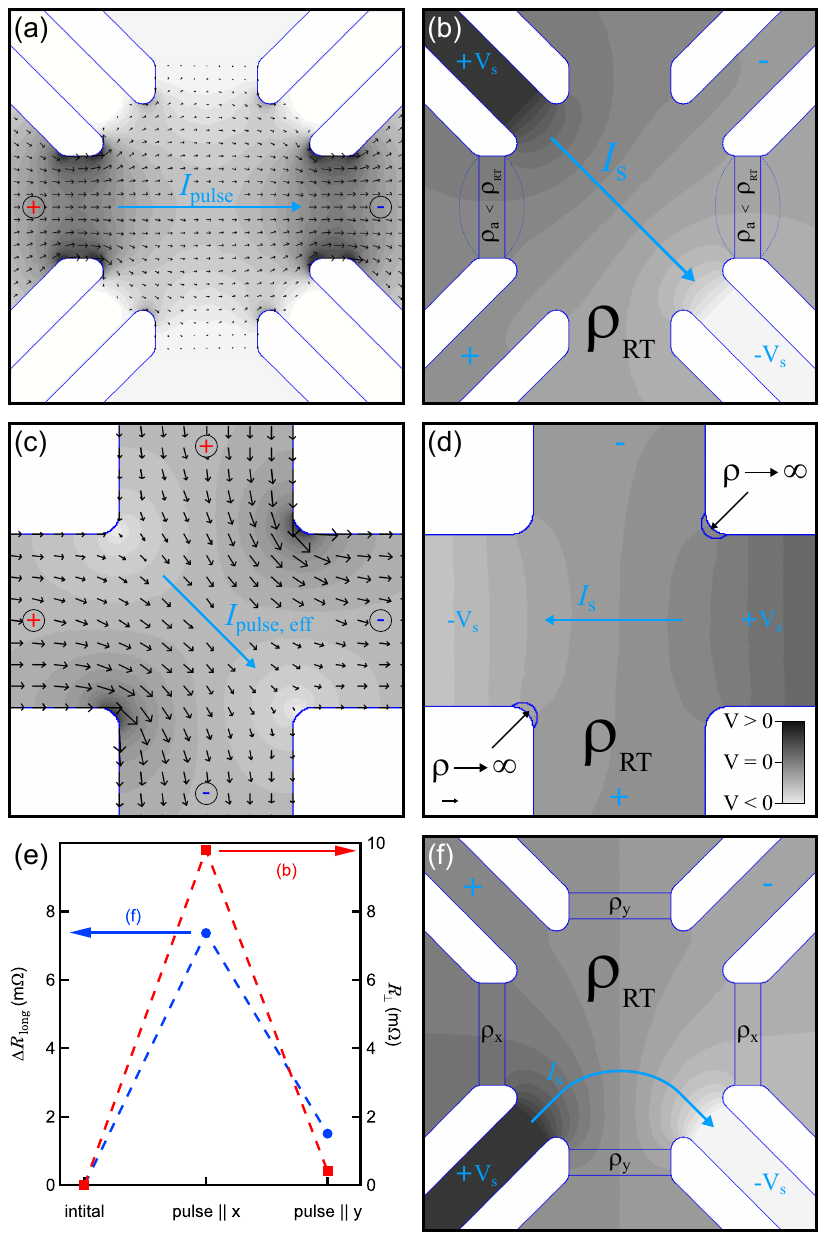}
	\caption[fig6]{\textbf{FEM simulation.}
		(a),~(c)~Current density distribution during a pulse in (a) star- and (c) cross-shaped devices with homogeneous resistivity. Arrows indicate the conventional current direction.
		(b),~(d)~Voltage landscape in devices with locally varying resistivity $\rho_\text{RT}$, $\rho_\text{a}$ and $\rho \rightarrow \infty$. 
		(e)~Simulation of the pseudo-longitudinal resistance change $\Delta R_\text{long}$ for the measurement geometry (f) and the $R_\perp$ resistance for the measurement geometry (b). Pulsing $\parallel x$ reduces $\rho_x < \rho_\text{RT}$ and breaks the symmetry of the device. Pulsing  $\parallel y$ restores the symmetry of the device with $\rho_y = \rho_x < \rho_\text{RT}$.
		(f)~Potential landscape in star-shaped devices using the pseudo-longitudinal measurement geometry.
	}
	\label{fig:6}
\end{figure}

We interpret our experimental findings on Nb films as a symmetry breaking on the (macroscopic) device level rather than on the microscopic level. Due to the Joule heating associated with the large current densities, a local variation of the film resistivity $\rho$ is created. This resistivity change $\Delta \rho$ is located in the constrictions of the device where $j$ is large compared to the rest of the device. In the following, we attempt to understand the experimental evidence presented before by a finite-element model. Fig.~\ref{fig:6}\,(a) shows the locally resolved current density in our eight-arm devices during a pulse. Additionally, we consider another pulse-current distribution in cross-shaped devices, see Fig.~\ref{fig:6}\,(c), which we discuss later (cf. Ref. \cite{Olejnik2017, Meinert2018}). 

The large current density in the constrictions of the device was shown to locally change the film crystallinity and to reduce the resistivity as a consequence. For the simplicity of the model, we consider local changes of the resistivity in rectangular areas within the constrictions as depicted in Fig.~\ref{fig:6}\,(b). In our example, the dimensions of the pulse line constrictions are about $2 \times 8 \, \text{\textmu m}^2$ and each constriction accounts for $\sim 3\%$ of the total pulse line resistance $R_{x\left[y\right]}$. Assuming that $\rho$ only changes within the constrictions, a change $\Delta R_x = R_x - R_{x,0}$ can be translated into a resistivity change within the constriction
\begin{align}\label{eq:rho_c}
\frac{\Delta \rho_\text{c}}{\rho} = \frac{R_x/R_{x,0} - 0.94}{0.06} - 1
\end{align}
with $R_{x,0}$ being the pulse line resistance before pulsing. From the resistance variation of the pulse line in Fig.~\ref{fig:3}\,(b) we calculate a maximum $\Delta \rho_\text{c} / \rho $ of about $-6\%$ (around burst index 495). Here, we measure $\Delta R_\perp = 17.7\,\text{m}\Omega$. To check whether $\Delta \rho_c$ can account for $\Delta R_\perp$, we perform FEM simulations using the software FEMM~v4.2 \cite{Baltzis2008_FEMM, Meeker_FEMM}. The resistivity of the film is set to $\rho_\text{RT}$ and the resistivity in the constrictions is $\rho_\text{a} = \rho_\text{RT} + \Delta \rho_c = 0.94\,\rho_\text{RT}$ (cf. Fig.~\ref{fig:6}\,(b)). We then apply $I_\text{s}$ and find that a transverse voltage arises due to the symmetry break about the current direction. The simulated transverse resistance is $R_\text{FEM} = 8.7\,\text{m}\Omega$.  $R_\text{FEM}$ critically depends on the exact choice of the shape of the area with $\rho_\text{a}$, e.g. if we make the area slightly convex (see dotted lines in Fig.~\ref{fig:6}\,(b)), the result increases to $R_\text{FEM}^\text{convex} = 22.3\,\text{m}\Omega$. In Fig.~\ref{fig:3}\,(a) we show full calculations with the model using rectangular constrictions, where the local resistivity variation is calculated on the basis of the change in pulse line resistance. We find that the observed transverse resistance is qualitatively well reproduced by the model, although quantitative descrepancies remain, which are probably due to the simplicity of the model. Thus, our conclusion is that the local change $\Delta \rho_c$ in the constrictions of our device contributes largely to the transverse voltage signals. In addition, the sign of the transverse voltage is the same in the FEM calculation and in the experimental lower-current density regime of the RT sample. This implies that the sign change of the transverse resistance in the experiment with the RT sample (cf. Fig.~\ref{fig:2}\,(a)) requires a local increase of the resistivity. This is consistent with Fig.~\ref{fig:3}\,(a) and (b), where the increase of pulse line resistance correlates with a sign reversal in the transverse voltage change. The ``saw-tooth''-like electrical response of Fig.~\ref{fig:2}\,(a) and (b) is explained by considering that for pulsing along one direction the mirror symmetry about $I_\mathrm{s}$ is gradually broken with every pulse, whereas pulsing along the orthogonal channel restores the mirror symmetry by also reducing/increasing the resistivity in the constrictions. We simulated this situation with the FEM model and show the result in Fig.~\ref{fig:6}\,(e) (right axis). After restoring the mirror symmetry by reducing the local resistivities along both $x$ and $y$ pulse lines, the transverse voltage drops back to zero within numerical accuracy.

Now we come back to the cross-shaped devices that were used in several previous studies \cite{Meinert2018, Olejnik2017, Baldrati2018, Moriyama2018}. In this configuration, the pulse voltage is applied to pairs of contacts as depicted in Fig.~\ref{fig:6}\,(c). Here, the maximum current density is found in the corners of the structure. The readout is done with a simple Hall configuration, see Fig.~\ref{fig:6}\,(d). As we lack detailed measurements to compare with, we just show the basic principle of how a transverse voltage arises in this type of device. With respect to the Hall configuration, the mirror symmetry is again broken by the 
local resistivity change in the corners and restored by pulsing in the orthogonal direction.
In the simulation, we consider $\rho \to \infty$ for simplicity in a small rounded area close to the corners (Fig.~\ref{fig:6}\,(d)). Clearly, this gives rise to a transverse voltage, which can be seen by the distorted equipotential lines. Quantifying the local resistivity changes in this type of device is probably more difficult than in the star-shaped devices. It does therefore seem advisable to preferably make use of star-shaped devices in order to quantify the pulse line resistances for estimating the associated transverse voltage with FEM simulations.

Finally, in some cases a readout configuration as depicted in Fig.~\ref{fig:6}\,(f) is used, which we call ``pseudo-longitudinal''. Following previous symmetry-derived arguments, one would not expect an influence of the local resistivity changes. However, our FEM simulation shows that even in this case a change of the pseudo-longitudinal voltage picked up between the top-left and top-right contacts is visible when the constrictions along the $x$ and $y$ directions change resistivity, cf. Fig.~\ref{fig:3}\,(e). Here, the reason is that the current-density distribution moves along the $y$-direction, such that more current leaks into the film area between the pickup lines. Due to symmetry, this is not fully reversible by orthogonal pulsing and should lead to a ``saw-tooth'' signal with a prominent drift given by the inverse sign of the local resistivity change. This drift is readily visible in our simulation in Fig.~\ref{fig:6}\,(e). We also note that signal drifts may be ubiquitous as soon as the breaking and restoration of the device symmetry are not perfect, independent of the actual device shape.

\section{Conclusion}

\begin{figure}[b]
	\centering
	\includegraphics{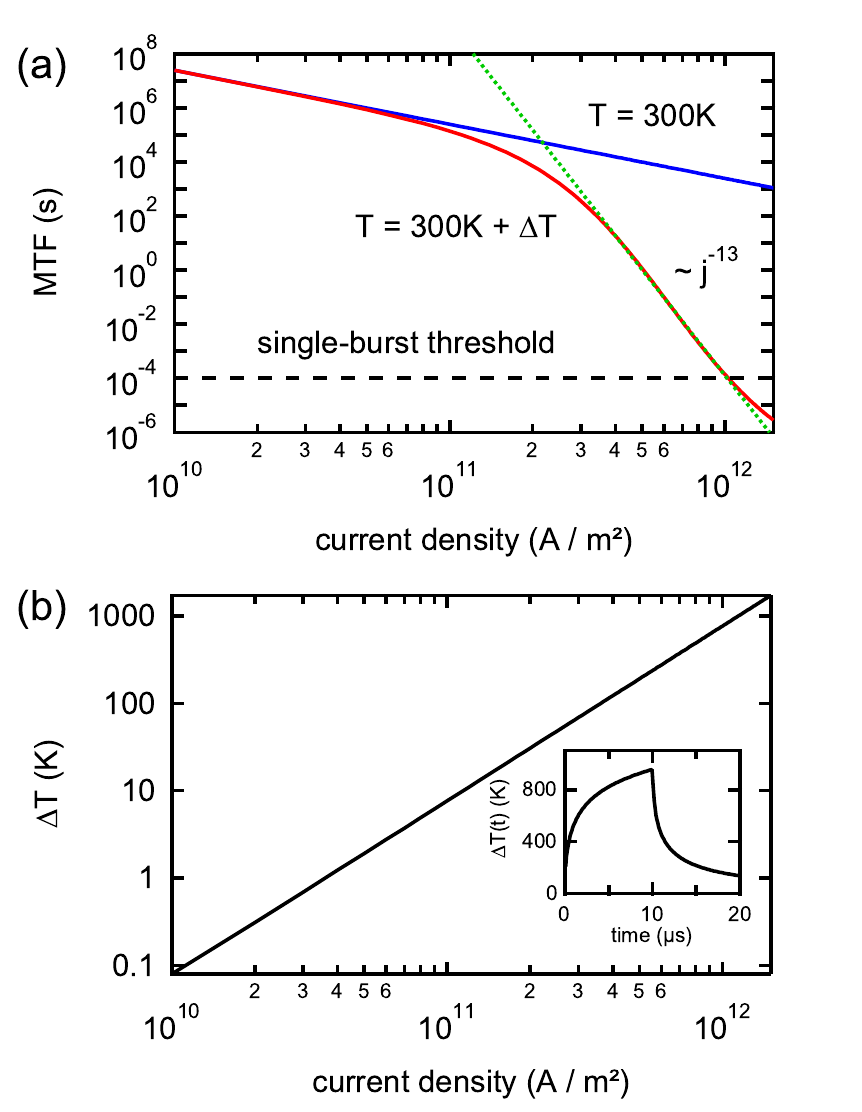}
	\caption[fig7]{\textbf{Black's equation.}
		(a) Mean time to failure (MTF) as a function of the current density in the constrictions of our devices. Two models are considered, one at fixed temperature of $300\,\mathrm{K}$ and one including the Joule heating contribution $\Delta T$. The dotted line represents a power-law fit to the high-current density regime.
		(b) $\Delta T$ as a function of the current density $j$ calculated with You's formula \cite{You2006}. The inset shows the time-dependence of the film temperature for a pulse of $\Delta t = 10$\,\textmu s and $j = 10^{12}\,\mathrm{A/m^2}$.
	}
	\label{fig:7}
\end{figure}

We can conclude that we see two different counteracting mechanisms: on the one hand, the resistivity reduction by crystallization. On the other hand, there is a mechanism that increases the resistivity and destroys the sample eventually. The annealing-like contribution should not be present in the HT sample which is already of good crystallinity. Consistently, we see only one sign of the transverse voltage change in the HT sample, which corresponds to a local increase of the resistivity according to the FEM simulation. The latter is supposedly related to electromigration, which transports material away from the constrictions, thereby creating voids in the material and increasing the local resistivity. The final destruction of the device goes along with a melting of the film and droplets of the material  moving in electron flow direction (cf. Fig.~\ref{fig:3}\,(f)). Hence, electromigration is probably the essential mechanism underlying the apparent resistivity increase and eventual breakdown of the device. We note that both the annealing-like component as well as the electromigration-like component are thermally activated and depend on individual energy barriers $Q$ and the temperature $T$ via a Boltzmann factor \cite{Jackson2004}. In particular, according to Black's equation \cite{Black1969}
\begin{align}\label{eq:black}
\mathrm{MTF} = \frac{A}{j^2} \exp{\frac{Q}{k_\mathrm{B}T}}
\end{align}
the mean-time-to-failure (MTF) is a strong function of current density and temperature, where the film temperature itself depends strongly on the current density. Here, $k_\mathrm{B}$ is the Boltzmann constant and $A$ is a scaling factor. Due to the high $j$ necessary to switch an AFM, the film temperature may rise by several hundred Kelvin during the pulses \cite{You2006, Meinert2018, Bodnar2018}. Thus, experiments performed close to the destruction threshold of the devices are particularly prone to the electromigration-like component of the electrical response and results should be taken with care. In Fig.~\ref{fig:7}, we evaluate Black's equation within two models, one calculated at fixed $T = 300\,\mathrm{K}$, and one calculated at an elevated temperature $T + \Delta T$, where $\Delta T$ is calculated as a pulse-averaged temperature with You's formula \cite{You2006} for a pulse width of 10\,\textmu s and typical parameters of the MgO substrate (see Ref. \cite{Meinert2018} for details). The parameters for Black's equation are chosen as $Q = 0.6\,\mathrm{eV}$ and $A = 2 \times 10^{17}\, \mathrm{sA^2/  m^4}$. While $Q$ is a typical value for grain-boundary self-diffusion \cite{Black1969}, $A$ is chosen to roughly match the experimental destruction threshold of the device. It can be seen in Fig.~\ref{fig:7}\,(a), that the Joule heating is largely responsible for the short MTF for $j > 10^{11}\,\mathrm{A/m^2}$. Fig.~\ref{fig:7}\,(b) demonstrates that for $j > 10^{12}\,\mathrm{A/m^2}$, the pulse-averaged film temperature is elevated by more than 700\,K and may reach even higher temperatures in the final stage of device breakdown, when the constriction becomes even more narrow due to the formation of cracks in the hot-spots. This is consistent with the observed droplets in the destroyed device, see Fig.~\ref{fig:3}\,(f). In the high-current density regime, Black's equation is often rewritten as $\mathrm{MTF} = A \, j^{-m} \exp{\left(Q/k_\mathrm{B}T\right)}$ to effectively take the Joule heating into account, where $m > 2$. Our numerical estimates suggest that $m\approx 13$ in this specific example, i.e. the MTF shows a very rapid decay with increasing $j$. Because of the similarities in the underlying thermal activation physics, we expect that similar dependencies on measurement temperature and current density may occur due to annealing/electromigration and N\'eel-order switching. Consequently, it is a difficult task to disentangle the various contributions by electrical experiments alone.

Using a chemically stable capping layer may reduce the surface contribution to the electromigration. However, the volume-contributions of the electromigration can not be reduced by cap layer engineering. As electromigration is a directed transport along the direction of electron flow, one may use single-cycle AC current pulses instead of DC pulses, similar to a previous experiment with CuMnAs \cite{Olejnik2018}. If the current-pulses are short enough, this should greatly reduce the directed transport, thereby increasing the device lifetime by orders of magnitude \cite{Liew1989}. As consequence, the resistivity-enhancing component of the transverse electrical response should be substantially reduced. However, the purely thermal, annealing-like contribution would still be present. This contribution should be relatively easy to remove by pulsed-current annealing the device with a larger current density before systematic investigations of the N\'eel-order switching are performed. In all cases, it is mandatory to carefully monitor the pulse line resistances and the repeatability of an electrical N\'eel-order switching experiment to avoid possible confusion of the true planar Hall signal due to switching and the resistive contributions.

As a final remark, we note that we performed similar experiments with films of Ti, V, Mo, and Ta with various degrees of crystal quality, which support our conclusions.

\section{Summary}

In summary, we investigated the resistive response of star-shaped devices often used in switching experiments with antiferromagnets. We used Nb thin films of different crystallinity and observed different responses in films with either polycrystalline or epitaxial growth. We reason that a change of the resistivity in the constrictions of the devices leads to a transverse voltage in planar Hall effect measurement geometry. From the pulse line resistance changes, we estimate the magnitude of resistivity change in the constrictions and perform FEM simulations that are in semi-quantitative agreement with the experimental data. The decrease and increase of the local resistivity is due to a crystallization and destruction of the sample, respectively. \textit{In operando} SEM imaging and \textmu XRD measurements support our interpretation. The crystallization or annealing effect is not visible in SEM images, whereas \textmu XRD provides evidence for the local crystallization in polycrystalline films. The sample destruction becomes significant in the transverse and pulse line resistance way before we were able to see any damage in SEM images. This effect resembles the planar Hall effect and is of nonmagnetic origin. Throughout this study, it was found to be irreversible and relaxation was not observed. 

For future electrical switching experiments it is thus advisable to regularly measure the pulse line resistances and to check the reproducibility of the switching in, e.g., current density sweeps. As the destructive contribution can be practically constant over a vast number of repeats, the reproducibility of the experiment alone without varying parameters and resistance monitoring does not yield sufficient evidence to exclude the discussed nonmagnetic contributions.

\subsection*{Acknowledgments}

This research used beamline 12.3.2 of the Advanced Light Source, which is a DOE Office of Science User Facility under contract no. DE-AC02-05CH11231. We further thank Christan Mehlhaff for conducting preliminary experiments.

\end{document}